\documentstyle[12pt]{article}
\begin{document}

\begin{center}
{\Large {\bf THE  IMAGINARY PART }} \\
\vspace{.7cm}
{\Large{\bf OF THE NUCLEON SELF-ENERGY  }}\\
\vspace{.7cm}
{\Large {\bf IN HOT NUCLEAR MATTER}}
\end{center}

\vspace{1cm}

\begin{center}
{\Large{L. Alvarez-Ruso$^a$, P. Fern\'andez de C\'ordoba$^b$}}
\end{center}
\begin{center}
{\Large{  and E. Oset$^a$}}
\end{center}

\vspace{1cm}

\begin{center}
{\small{\it $^a$Departamento de F\'{\i}sica Te\'orica and IFIC\\
Centro Mixto Universidad de Valencia-CSIC\\
Valencia, Spain}}
\end{center}
\begin{center}
{\small{\it $^b$Departamento de  Matem\'atica Aplicada\\
Universidad Polit\'ecnica de Valencia\\
Valencia, Spain}}
\end{center}

\vspace{3cm}

\begin{abstract}
{\small{ A semiphenomenological approach to the nucleon self-energy in nuclear
matter at finite temperatures is followed. It combines elements of Thermo 
Field 
Dynamics for the treatment of finite  temperature with a model for the
self-energy, which evaluates the second order diagrams taking the
needed dynamics of the NN interaction  from experiment.
The approach proved to be accurate at zero temperature to reproduce 
$Im \Sigma$ and other properties of  nucleons in matter. In the present case 
we apply it to determine $Im \Sigma$ at finite temperatures. An
effective NN cross section is deduced which can be easily used
in analyses of heavy ion reactions.}}
\end{abstract}

\newpage

\section{Introduction}

The imaginary part of the nucleon self-energy $\Sigma$
 has been the subject of
intense research in the past \cite{1}. However, these studies have
been done at zero temperature, of relevance for physical processes involving
nucleon-nucleus scattering. On the other hand, in heavy ion collisions, 
through multiple collisions of the nucleons, one reaches conditions roughly
similar to those of a thermal bath  at finite temperature \cite{2}. The
imaginary part  of the nucleon self-energy 
in such a bath is an important magnitude
which governs processes of nucleon emission, particle production and, in
principle, most of the nuclear processes occurring during
 heavy ion collisions, including the rate of thermalization.

The many body field theoretical treatment at finite temperature becomes,
however, technically more involved than at $T = 0$. At the root of it
lies the fact that particle annihilation operators do not
give a vanishing result when applied to the ground state of the system,
which is now a thermal distribution. This does not allow one to use the
Wick theorem to generate the Feynman diagrammatic perturbation expansion,
as we know for many body systems at $T = 0$, or in ordinary Quantum Field
Theory in the absence of a medium. Yet, even if more complicated, the 
methods to deal with it are available. The traditional approach has been
the one of the imaginary time formalism \cite{3,4}, although lately
the new method in real time formalism known as Thermo Field Dynamics
\cite{5,6} is proving quite efficient and becomes more and more widely
used. A recent review on Thermo Field Dynamics with some applications to 
nuclear matter looking at collective modes
and delta propagation in matter is presented in ref. \cite{7}. A covariant 
formalism at finite temperature unifying the good features of the two
formalisms is presented in ref. \cite{8}, with some application to
nuclear matter within the relativistic Walecka model \cite{9}.

Detailed analyses of nucleon properties along the lines of ref. \cite{1}
are available at finite, but small temperatures $T \leq  10 \, MeV$
\cite{10,11}. Neither of the two approaches mentioned above is
followed, but instead the smallness of the temperature justifies some
approximations by means of which one finally follows the steps at
$T = 0$  substituting  the Pauli distribution $ n (\vec{k}) \,
( 1$ for $ |\vec{k} | < k_{F} , \, 0 $ for $|\vec{k}| >
k_{F})$ by the thermal distribution

\begin{equation}
n (k^0) = [1 + exp (\frac{k^0 - \mu}{k_B T})]^{-1} 
\end{equation}

\noindent
with $k_B$ the Boltzmann constant, $k^0 = \varepsilon (\vec{k})$ 
the nucleon energy
and $\mu$ the chemical potential. The results of \cite{10} are
improved in \cite{11} by considering the correlation diagrams which lead
to a large contribution to $Im \Sigma$ at momenta below the Fermi momentum.
Similar steps are followed in ref. \cite{12} to deal with the propagation
of the $\Delta$ at finite temperatures.

The appeal of Thermo Field Dynamics is that one can continue to use
Wick's theorem and the Feynman diagrammatic approach as at $T = 0$. The
price is the introduction of an auxiliary field by means of which one
constructs a field doublet; the propagators become 2 $\times$
 2 matrices and the
Feynman rules are now algebraic operational rules in the space of
2 $\times$ 2 matrices.

In the present work we shall evaluate the imaginary part of the nucleon
self-energy at finite temperature. In doing so we shall be following
closely the steps of ref. \cite{13}, where $Im \Sigma$ was evaluated for
nuclear matter at $T = 0$. The approach used in \cite{13} was 
semiphenomenological, much in line with usual approximations done in the
treatment of heavy ion collisions where the results obtained here are
bound to be of relevance. The approach of \cite{13} evaluated
correctly the second order nucleon self-energy diagrams but bypassed the
use of an explicit NN potential. Instead, it used the fact that the
sum of ladder diagrams in the low density limit provides the NN scattering
t matrix, and its modulus squared, which appears in the evaluation
of $Im \Sigma$, is related to the NN cross section. Hence the experimental
cross section was used as input and the low density theorem \cite{14,15}
was automatically fulfilled. This theorem states that, as the nuclear density
goes to zero, one has

\begin{equation}
Im \Sigma_{\rho \rightarrow 0} = - \frac{1}{2} \  \sigma_{tot} \  v \  
 \rho
\end{equation}

\noindent
where $v$ is the velocity of the nucleon in the rest frame of the Fermi
sea and $\sigma_{tot}$ the total NN cross section. Long range correlations
were also considered by means of an RPA sum with a phenomenological 
particle-hole
interaction acting in the spin-isospin transverse channel. The approach 
proved to be numerically quite successful by comparing the results with
those of elaborate and time consuming many body approaches like those
of refs. \cite{16,17}. Spectral functions and occupation numbers
were also evaluated in \cite{13} and were very similar to 
those obtained in other successful many body approaches with special
emphasis in selfconsistency \cite{18}.

As with respect to the finite temperature treatment, we use the Thermo
Field Dynamics approach by following the formalism, normalization
and rules of ref. \cite{19}, where a comprehensive and practical extract
of this method is presented.

\section{Finite temperature formalism for the nucleon self-energy}

By following ref. \cite{19} we have the thermal doublet for the nucleon
field

\begin{equation}
\psi^{(a)} (x) \equiv 
\left\{
\begin{array}{c}
\psi (x)\\
i \  ^t\tilde{\psi}^\dagger
\end{array}
\right\} 
\end{equation}

\noindent
where $\psi (x)$ is the ordinary nucleon field and $\tilde{\psi}$ a
support field for the formalism ($t$ means the transposition with respect to
the spinor index and $\dagger$ stands for conjugate). The 
thermal propagator is now defined as

\begin{equation}
i G^{(a,b)} (x_{1}, x_{2}) = <0(\beta) | T [\psi^{(a)} (x_{1}) \bar{\psi}^{(b)}
(x_{2})] | 0 (\beta) >
\end{equation}

\noindent
where $| 0 (\beta)>$ is the state transformed from the vacuum by a
Bogoliubov unitary transformation, and which has the  property that 
the expectation value of an operator in this state is
equivalent to its thermal average with the distribution
of eq. (1). Hence the component of the Green's function, $G^{(11)}$, is
the proper thermal average of the ordinary Green's function.

The thermal free propagator, to be used in perturbation theory  in
the nonrelativistic approximation which we shall follow here, is
given by

\begin{equation}
G^{0\, (ab)} (p) = G^{0\, (ab)}_{F} (p) + G^{0\, (ab)}_D (p)
\end{equation}

\noindent
with

\begin{equation}
G^{0\, (ab)}_F (p) = 
\left(
\begin{array}{cc}
G^0_F (p) & 0\\
0 & G^{0 *}_F (p)
\end{array} \right)
\, ; \, G^0_F (p) = 
\frac{1}{p^0 - \varepsilon (\vec{p}) + i\epsilon}
\end{equation}

\vspace{0.3cm}

\begin{equation}
G_D^{0\, (ab)} (p) = 2 \pi i \delta (p^0 - \varepsilon (\vec{p}))
\left( 
\begin{array}{cc}
sin^2 \theta_{p^0} & \frac{1}{2} sin 2 \theta_{p^0}\\[2ex]
\frac{1}{2} sin 2 \theta_{p^0} & - sin^2 \theta_{p^0}
\end{array} \right)
\end{equation}

\vspace{0.3cm}

\begin{equation}
cos \theta_{p^0} = \frac{1}{(1 + e^{-x})^{1/2}} \, ; \, 
sin \theta_{p^0} = \frac{e^{- x/2}}{(1 + e^{-x})^{1/2}} \, ;
x = \frac{p^0 - \mu}{k_B T}
\end{equation}

\vspace{0.3cm}

\begin{equation}
sin^2 \theta_{p^0} = \frac{1}{e^x + 1} = n (p^0)
\end{equation}

\noindent This 
Green's function is in addition diagonal in spin but we omit the
spin indices for simplicity.

An alternative way of writing this propagator which is also suited to write the
exact propagator is

\begin{equation}
G^{0 (ab)} (p) = \left[ U_F (p^0) \left( 
\begin{array}{cc}
G^0_F (p) & 0\\
0 & G^{0*}_F (p)
\end{array}
\right)
U_F^{-1} (p^0) \right]^{(ab)}
\end{equation}

\noindent
with

\begin{equation}
U_F (p^0) = \left[ 
\begin{array}{cc}
cos \theta_{p^0} & sin \theta_{p^0}\\
- sin \theta_{p^0} & cos \theta_{p^0}
\end{array} \right]
\end{equation}

\noindent
and the exact propagator can be cast as

\begin{equation}
G^{(ab)} (p) = U_F (p^0) 
\left(
\begin{array}{cc}
[p^0 - \varepsilon(\vec{p}) - \bar{\Sigma} (p)]^{-1} & 0\\
0 & [p^0 - \varepsilon (\vec{p}) - \bar{\Sigma}^* (p)]^{-1}
\end{array} \right)
U^{-1}_F (p^0)
\end{equation}

\noindent
with $\varepsilon (\vec{p})$ the kinetic energy of the particle and

\begin{equation}
\begin{array}{ll}
Re \bar{\Sigma} (p) & = Re \Sigma^{(11)} (p)\\
Im \bar{\Sigma} (p) & =  Im \Sigma^{(11)} (p) / cos 2 \theta_{p^0}
\end{array}
\end{equation}

In particular  the $G^{0 (11)} (p)$
component has an intuitive form given by

\begin{equation}
G^{0 (11)} (p) = \frac{1- sin^2 \theta_{p^0}}{p^0 - \varepsilon(\vec{p})
 - \bar{\Sigma}} + \frac{sin^2 \theta_{p^0}}{p^0 - \varepsilon(\vec{p}) 
 - \bar{\Sigma}^*} 
\end{equation}

\noindent which in the limit of $T = 0$ reproduces the standard form of the
nucleon propagator in a Fermi sea.

Hence in order to obtain the self-energy $\bar{\Sigma} (p)$ 
which renormalizes the nucleon propagator, only the thermal component
$\Sigma^{(11)}$ needs to be evaluated.
 In the next section we show the approximation scheme that we
follow to evaluate this magnitude.

\section{Semiphenomenological model for $\Sigma$}

This section follows closely ref. \cite{13}. 
The generic Feynman diagram which we evaluate is the one in fig. 1, where
the nucleon propagator in each of the baryonic lines is given by
eq. (5).  Note that in the limit of
$T = 0$, and with the conventional separation of particles and holes, the
usual polarization (fig. 2a) and correlation (fig. 2b) graphs which 
lead to $Im \Sigma$ \cite{11} are automatically generated (together 
with other two graphs with the interaction lines crossed which do not
 contribute  to $Im \Sigma$).

In fig. 1  the indices
 a,b,c,d in the vertices are thermal indices. We are 
interested in $\Sigma^{(11)}$ and hence $a = b = 1$. Assuming for the 
moment the  interaction lines to correspond to meson exchange, and
considering
 also that no such mesons are present in the 
ground state of our many body 
fermionic system, the meson propagator would be diagonal in the thermal
indices and hence $c =  d = 1$.

We thus must evaluate the polarization function $\Pi^{(11)} (q)$

\begin{equation}
\Pi^{(11)} (q) = - 4 i \int \frac{d^4 p}{(2 \pi)^4} \ \ G^{0\,(11)}(p) 
\ \ G^{0 \, (11)}
(p + q)
\end{equation}

\noindent
where the factor 4 takes into account the sum over spin and isospin.

Once again in the limit of $T = 0$, this polarization
would account for the two diagrams in fig. 3, which are those accounted
for by means of the Lindhard function \cite{4}.

By  using $G^{0 \, (a,b)}$ from eqs. (6,7)
the $p^0$ integration can be easily performed and one obtains

\begin{equation}
\Pi^{(11)} (q) = 4 \int \frac{d^3 p}{(2 \pi)^3} \left\{
\frac{sin^2 \theta_{\varepsilon (\vec{p})} cos^2 \theta_{
\varepsilon (\vec{p} + \vec{q})}}{q^0 + \varepsilon (\vec{p})
- \varepsilon (\vec{p} + \vec{q}) + i \epsilon}  + \frac{
sin^2 \theta_{\varepsilon (\vec{p} + \vec{q})} cos^2 \theta_{
\varepsilon (\vec{p})}}{- q^0 - \varepsilon (\vec{p}) +
\varepsilon (\vec{p} + \vec{q}) + i \epsilon}  \right\}
\end{equation}

\noindent
Next we evaluate $\Sigma^{(11)}$ corresponding to fig. 1

\begin{equation}
\Sigma^{(11)}(k) = i \int \frac{d^4 q}{(2 \pi)^4} \ \
 V^2 (q) \ \  \Pi^{(11)} (q) \ \
G^{0 \, (11)} (k - q)
\end{equation}

\noindent
where $V (q)$ would take into account the interaction due to the 
hypothetical meson exchange.

Here again we follow the steps of ref. \cite{13} and sum the ladder
diagrams which would replace $V (q)$ by the scattering $t$ matrix.
(Note that medium corrections to $t$ which would appear in the
medium $G$-matrix are taken explicitly into account to second order
with the structure of the diagram). 
We shall continue
to use the same $t$ matrix here. The studies of refs. \cite{10,11,20}
show indeed little dependence of the effective interaction on the
temperature.

In order to evaluate $Im \Sigma^{11)}$ from eq. (17) a Wick rotation
was made in ref. \cite{13} which allows one to express $Im \Sigma^{(11)}$
in terms of $Im \Pi^{(11)}$. This is however not possible here 
because $\Pi^{(11)}$ from eq. (16) has overlapping cuts in the upper
and lower half planes of the complex plane (unlike at $T = 0$
where the cuts are confined to the second and fourth quadrant). However,
an explicit evaluation of $Im \Sigma^{(11)}$ is possible by first 
performing the $q^0$ integral in eq. (17) and  then evaluating the 
imaginary part, with the result

\begin{equation}
\begin{array}{l}
Im \Sigma^{(11)} (k) = - 4 \pi \int \frac{d^3 q}{(2 \pi)^3}
\int \frac{d^3 p}{(2 \pi)^3} |t|^2 \delta (k^0 + \varepsilon (
\vec{p}) - \varepsilon (\vec{k} - \vec{q}) - \varepsilon (\vec{p}
+ \vec{q}) ) \\
\cdot \left\{ cos^2 \theta_{\varepsilon (\vec{k} - \vec{q})}
cos^2 \theta_{\varepsilon (\vec{p}+ \vec{q})} 
sin^2 \theta_{\varepsilon (\vec{p})}
- sin^2 \theta_{\varepsilon (\vec{k} - \vec{q})}
sin^2 \theta_{\varepsilon (\vec{p} + \vec{q})} cos^2 \theta_{
\varepsilon (\vec{p})} \right\}
\end{array}
\end{equation}

The spin-isospin averaged value of $|t|^2$ assumed in eq. (18)
is then replaced by $4 \pi \sigma_{NN} /M^2$, where M is the nucleon
mass and $\sigma_{NN}$ the spin-isospin averaged NN elastic cross section.
Since pion production is not explicitly taken into account, this
restricts us below the pion production threshold. The final step in
ref. \cite{13} is to consider the polarization or RPA 
corrections to the interaction. 

The consideration of the polarization was an important ingredient in
ref. \cite{13}, which reduced $Im \Sigma$  particularly at small energies,
and provided results similar to those found in refs. \cite{16,17}.
We implement it here too. For this purpose we need to evaluate $\Pi^{(11)}
(q)$, both the real and imaginary part, which cares about $ph$ excitation,
and $\Pi_\Delta ^{(11)} (q)$, the equivalent term accounting for
$\Delta h$ excitation. At $T = 0$ these quantities are the familiar
Lindhard functions $U_N (q), U_\Delta (q)$, respectively, used in
ref. \cite{13}. 

The real part of $U_N (q)$, unlike $Im U_N (q)$, is not affected by
Pauli blocking \cite{4}, hence finite temperature, which modifies
occupation numbers, has not much of a consequence in the
change of $Re U_N (q)$. On the other hand there is no Pauli blocking in the
$\Delta h$ excitation since we do not have a Fermi sea of $\Delta$'s.
For these reasons we keep
$Re \Pi^{(11)}$ and $\Pi^{(11)}_\Delta$ at finite temperatures equal
to $Re U_N (q), U_\Delta (q)$ at zero temperature. However, we 
evaluate  $Im \Pi^{(11)} (q)$ from eq. (16) at finite temperature.
The reason is that keeping $Im \Pi^{(11)} \neq 0$ is
important in order to avoid singularities coming from poles of zero
sound ($q^0$ proportional to $|\vec{q}|$ at small energies)  which are
strongly dumped  at finite $T$.

The expression for $Im \Pi^{(11)}$ obtained from eq. (16) is given
by

\begin{equation}
\begin{array}{ll}
Im \Pi^{(11)} (q^0, q) = - \frac{1}{\pi} & \int^\infty_{p_{min}} dp
\frac{mp}{q} \{ sin^2 \theta_{\varepsilon (\vec{p})} + sin^2 \theta_{
\varepsilon (\vec{p}) + q^0}\\[3ex]
& - 2 sin^2 \theta_{\varepsilon (\vec{p})} sin^2 \theta_{\varepsilon (\vec{p})
+ q^0} \}
\end{array}
\end{equation}

\noindent
where

\begin{equation}
p_{min} = \frac{m}{p} \; | q^0  - \frac{\vec{q}\,^2}{2m}|
\end{equation}

The polarization correction  replaces the interaction by the
induced interaction \cite{21} (see eq. (15) of ref. \cite{13}). Furthermore,
we can perform some trivial integrals and eliminate the $\delta$
function with the final result

\begin{equation}
\begin{array}{l}
Im \Sigma^{(11)} (k) = - \frac{\sigma_{NN}}{M \pi^2} \int_0^\infty
q d q \int_{-1}^{1} d cos \theta \int_0^\infty p dp \,\Theta
(1 - A^2)\\
\cdot \frac{1}{|1 - V_t (q) U (q)|^2}|_{q^0 = k^0 - \varepsilon (\vec{k} - 
\vec{q})}\\
\cdot \left\{ cos^2 \theta_{\varepsilon (\vec{k} - \vec{q})}
cos^2 \theta_{\varepsilon (\vec{p} + \vec{q})} 
sin^2 \theta_{\varepsilon (\vec{p})}
- sin^2 \theta_{\varepsilon (\vec{k} - \vec{q})}
sin^2 \theta_{\varepsilon (\vec{p} + \vec{q})} cos^2 \theta_{\varepsilon
(\vec{p})} \right\}
\end{array}
\end{equation}

\noindent
where $ V_t (q)$ is the transverse part of the spin-isospin interaction
and $U (q) = \Pi^{(11)} (q) +
\Pi^{(11)}_\Delta (q)$. The arguments leading to this 
modifications
and expressions for $V_t (q)$ and $U_N (q)$, $U_{\Delta} (q)$ 
can be found in ref. \cite{13}
and we do not repeat them here.
Furthermore the angle in the integral over $cos \theta$ in eq. (21)
is the angle between $\vec{q}$ and $\vec{k}$. The magnitude
$A$ in the argument of the step function is given by

\begin{equation}
A = \frac{M}{pq} \left\{ k^0 - \frac{k^2}{2M} -
\frac{q^2}{M} + \frac{k q cos \theta}{M} \right\}
\end{equation}

\noindent
with $k, q$ the modulus of $\vec{k}$ and $\vec{q}$ respectively.

The value of the chemical potential $\mu$ as a function of the density and 
$T$ is obtained, as usually done, by the normalization condition

\begin{equation}
\rho = 4 \int \frac{d^3 k}{(2 \pi)^3} \frac{1}{1 + exp [(\varepsilon (k) -
 \mu) / k_B T]}
\end{equation}

Eq. (21) provides $Im \Sigma^{(11)}$ as a function of $k^{0},\vec{k}$.
In the next section we show the results which we obtain for this quantity.

\section{Results and discussion}

In fig. 4 we show the results of $- Im \bar{\Sigma}$ at $T = 0$ for two
densities, $\rho_0$ $(0.17 \, fm^{-3})$ and $\rho_0 /2$, obtained with the 
present formalism in the limit of $T = 0$. The results agree with
those in ref. \cite{13} calculated with the $T = 0$ formalism and also
with those of the microscopic approach of ref. \cite{16}. Note
that since only kinetic energies are used  as input to evaluate
 $\Sigma^{(11)}$, the value of $\mu$ is referred to an origin of
energies at $|\vec{k}| = 0$. We  are not interested in 
$Re \Sigma$ and to overcome the arbitrary origin of energies we plot
the magnitudes in terms of $\omega - \mu$ ( $\omega \equiv k^{0}$ ). 
In fig. 4, $|\vec{k}|$
 is taken at the value $\sqrt{2 M \omega}$. This justifies small
 differences with ref. \cite{13} where the value of $|\vec{k}|$
 satisfying the dispersion relation between $|\vec{k}|$ and
$\omega$ was taken. In fig. 4 we observe
the typical features that $Im{\bar{\Sigma}}$ is proportional to 
$(\omega - \mu)^2$. In the calculations we find
that $Im \Sigma^{(11)}$ changes sign at $\omega = \mu$, with 
$Im\Sigma^{(11)} < 0$ for $\omega > \mu$. 
In this case 

\begin{equation}
\begin{array}{ll}
cos 2 \theta_{k^0} & = 1 - 2 sin^2 \theta_{ k^0}\\
 & = 1 - 2 n (k^0)_{T = 0} = \left\{
\begin{array}{cc}
- 1 & \omega < \mu\\
1 & \omega > \mu
\end{array} \right.
\end{array}
\end{equation}

\noindent
and then

\begin{equation}
Im \bar{\Sigma} = \frac{Im \Sigma^{11}}{cos^2 \theta_{k^0}} = - 
| Im \Sigma^{11}|
\end{equation}

The Green's function $G^{(11)}$, by using eq. (12), will
then be

$$
\frac{\Theta (\omega - \mu)}{k^0 - \varepsilon (\vec{k}) - Re \Sigma^{(11)}
(k) + i |Im \Sigma^{(11)}|} + \frac{\Theta (\mu - \omega)}{k^0
- \varepsilon (\vec{k}) - Re \Sigma^{(11)} (k) - i |Im \Sigma^{(11)}|}
$$

\begin{equation}
\equiv \frac{1}{k^0 - \varepsilon (\vec{k}) - \Sigma^{(11)}}
\end{equation}

\noindent
as it should be.

In fig. 5 we plot $Im \Sigma^{(11)}$ at $\rho = \rho_0$ as a function
of $\omega - \mu$, with $|\vec{k}|=\sqrt{2 M \omega}$ and $\mu$ calculated
from eq. (23). 
As can be seen in the figure, $Im \Sigma^{(11)}$ is always zero at
$\omega = \mu$.
However, $Im \bar{\Sigma} (\mu, k)$ is different from zero at finite
temperatures, contrary to the situation at $T = 0$ where it
is zero. In order to envisage this we see that the evaluation of
$Im \bar{\Sigma}$ from eq. (13) involves a fraction of the type
$0/0$ which we determine using l'H\^{o}pital rule and find

\begin{equation}
Im \bar{\Sigma} (\mu,k) = \lim_{k^0 \rightarrow \mu} \frac
{Im \Sigma^{(11)} (k^0,k)}{1- 2 n(k^0)} = 2 k_B T \frac{d}
{d k^0} Im \Sigma^{(11)} (k^0,k) |_{k^0=\mu}
\end{equation}

\noindent
We can  see that $Im \bar{\Sigma} (\mu, k)$
 vanishes at $T = 0$, as
we already said.

In fig. 6,7,8 we show  the results  for $- Im \bar{\Sigma}$
 as a function of $\omega - \mu$ for different temperatures, calculated
for three different densities, $\rho_0/2,\rho_0$ and $2 \rho_0$.
We can appreciate that as $T$ increases  $- Im \bar{\Sigma}$
also increases in all the range of energies calculated there.
 We should note that in evaluating
 $Im \bar{\Sigma}$, the factor $cos 2 \theta_{k^0}$ appearing
in the denominator of eq. (13) is very important
and makes  $Im \bar{\Sigma} \ne 0$ at $T \ne 0$ for all the
range of energies, while  $Im \Sigma^{(11)}$
passed through zero.
This is a genuine temperature dependent property which would
be lost if a $T = 0$ formalism, changing the Fermi
distribution by the thermal distribution of eq. (1), were used.

At this point it is interesting to compare our results for $Im \bar{\Sigma}$
with those which we would obtain using standard formulae of Fermi-liquid 
theory \cite{22}. The formula used there in our notation for 
$Im \bar{\Sigma}$, removing the cut off in the integral, 
would be given by eq. (18) changing the minus sign in 
the curled bracket ($cos^2$ and $sin^2$ terms) by a positive sign.
Instead our formula for $Im \bar{\Sigma}$ uses eq. (18) with a minus
sign (which provides $Im \Sigma^{(11)}$) and then we divide by 
$cos 2 \theta_{po}$ (as shown in eq. (13)) in order to obtain 
$Im \bar{\Sigma}$.
The same prescription would be taken if one uses instead eq. (21) which 
incorporates the effects of polarization.

It is easy to see that at
$T = 0$ both formulae are identical. Indeed for $\omega > \mu$ only
the first term in the curled bracket of eq. (18) (or (21)) (the one with
$cos^2 cos^2 sin^2$) contributes, while for $\omega < \mu$ only the second
term in the bracket (the one with $sin^2 sin^2 cos^2$) contributes.
Then when dividing $Im \Sigma^{(11)}$ by $cos 2 \theta_{p^0}$, given by eq.
(24), we obtain a formula for $Im \bar{\Sigma}$ given by eq. (18) where
the minus sign in the curled bracket is changed to a positive sign, 
exactly the formula used in Fermi-liquid theory \cite{22}.

The identity of the two formulae holds, however, at any temperature. This 
can be seen analytically using eqs. (8) for $sin \theta, cos \theta$ and $x$, 
 hence

\begin{eqnarray}
& & \frac{(cos^2 \theta_{\varepsilon_{1}} cos^2 \theta_{\varepsilon_{2}} 
sin^2 \theta_{\varepsilon_{3}} - sin^2 \theta_{\varepsilon_{1}}
sin^2 \theta_{\varepsilon_{2}} cos^2 \theta_{\varepsilon_{3}})}
{cos 2 \theta_{k^0}} 
\delta (k^0 + \varepsilon_{3} - \varepsilon_{1} - \varepsilon_{2} )
\nonumber
\\
&=& \frac{1}{k_{B} T} \frac{e^{x_{3}} [ e^{( x_{1}+x_{2}-x_{3} )} - 1 ]}
{( 1+e^{x_{1}} ) ( 1+e^{x_{2}} ) ( 1+e^{x_{3}} )}
\frac{e^{x_{k^0}}+1}{e^{x_{k^0}}-1} \delta (x_{k^0}+x_{3}-x_{1}-x_{2})
\nonumber
\\
&=& \frac{1}{k_{B} T} \frac{e^{x_{1}} e^{x_{2}} + e^{x_{3}}}
{( 1+e^{x_{1}} ) ( 1+e^{x_{2}} ) ( 1+e^{x_{3}} )} 
\delta (x_{k^0}+x_{3}-x_{1}-x_{2})
\nonumber
\\
&=& (cos^2 \theta_{\varepsilon_{1}} cos^2 \theta_{\varepsilon_{2}} 
sin^2 \theta_{\varepsilon_{3}} + sin^2 \theta_{\varepsilon_{1}}
sin^2 \theta_{\varepsilon_{2}} cos^2 \theta_{\varepsilon_{3}}) 
\delta (k^0 + \varepsilon_{3} - \varepsilon_{1} - \varepsilon_{2})
\nonumber
\\
& &
\end{eqnarray}

\noindent
where the constraints of the $\delta$-function have been used in the 
second step.
This is an interesting finding which stresses the
value of the results obtained in Fermi-liquid theory based on
the concept of quasiparticles, by comparison to a method in
principle more microscopic, like the one used here.

Next we would like to extract a practical
magnitude from these results which can be used in calculations
of heavy ion collisions. Recall
that in the semiclassical approach one has

\begin{equation}
 Im \bar{\Sigma} = -\frac{1}{2} \ \sigma_{NN} \ v \ \rho
\end{equation}

\noindent
One can then define an effective NN nucleon-nucleon cross section
at different $T$ and $\rho$ by means of

\begin{equation}
\sigma_{NN}^{eff}  =  - 2 \  \frac{ Im \bar{\Sigma}}{ v \rho}
\end{equation}

\noindent 
as done in ref. \cite{23}, where $\sigma_{NN}^{eff} \  \rho$
defines a probability of collision per unit
length for the nucleon.
In order to facilitate the use
of this magnitude we have parameterized  $v
   \sigma_{NN}^{eff}$, with $v=|\vec{k}|/M$
for the different densities and temperatures evaluated here.
We take the following functional form 

\begin{equation}
v \sigma_{NN}^{eff} = \sum_{n=0}^{4} a_n(\rho,T) \ \omega^n
\end{equation}

\noindent 
where $\omega$
is the nucleon kinetic energy, $\vec{k}^2/2 M$. The fit is valid
for values of $\omega > \mu$ in figures 6,7,8.
The coefficients $a_n(\rho,T)$ are given in tables I,II,III.
One can obtain  $v \sigma_{NN}^{eff} $ for intermediate
values of $\rho$ and $T$  interpolating
between the values of $v \sigma_{NN}^{eff} $ given by
eq. (29). 

\section{Conclusions}

	We have used a model to evaluate $Im \bar{\Sigma}$ for a nucleon in
nuclear matter at finite temperatures, which combines the formalism of 
Thermo Field Dynamics with empirical magnitudes of the NN interaction.
This model at $T = 0$ coincides with a semiphenomenological
approach studied earlier, which proved rather successful in reproducing
nucleon properties in nuclear matter obtained with more microscopical
approaches.

	We have obtained $Im \bar{\Sigma}$ for different values of the 
nuclear
density and several temperatures. One of the interesting findings is that
$Im \bar{\Sigma}$ grows steadily with the temperature. The changes produced by
the temperature are more striking at energies around the chemical potential
where $ Im \bar{\Sigma}$ is zero at $T = 0$ and  takes finite values at 
$T \ne 0$.
 
	We found that the genuine effects of the temperature, given naturally
in the formalism of Thermo Field Dynamics, were  important, and the 
differences with respect to simple calculations, where $n (\vec{k})$ 
at $T = 0$ is replaced by the thermal distribution, can be  appreciable.
	In order to facilitate the use of the results obtained here, we have
parameterized them by means of easy analytical formulae. The parameterization
is given for an effective NN cross section in the medium, such that 
$\sigma^{eff}_{NN} \  \rho$ gives the probability of collision per unit length
for a nucleon in the nuclear medium.
With the given formulae  one can easily interpolate the results and obtain
this magnitude for different densities and temperatures. These results should 
be useful in the analysis of heavy ion reactions.

\section*{Acknowledgements}

We would like to thank Liang-gang Liu and Igor Tkachenko for useful
discussions. This work has been partly supported by CICYT
contract number AEN 93-1205.

\newpage
\section{Table Captions}
\bigskip
\parindent 0 cm

{\bf Table I } Parameters of eq. (28) to evaluate $v \sigma^{eff}_{NN}$ at 
$\rho = \rho_{0}/2$. The parameters $a_{n}$ have dimensions such that, with 
$\omega$ given in $MeV$, the results for $v \sigma^{eff}_{NN}$ are given in 
$mb$.
\medskip

{\bf Table II } Same as table I for $\rho = \rho_{0}$.
\medskip

{\bf Table III } Same as table I for $\rho = 2 \rho_{0}$.

\newpage
\section{Figure Captions}
\bigskip
\parindent 0 cm

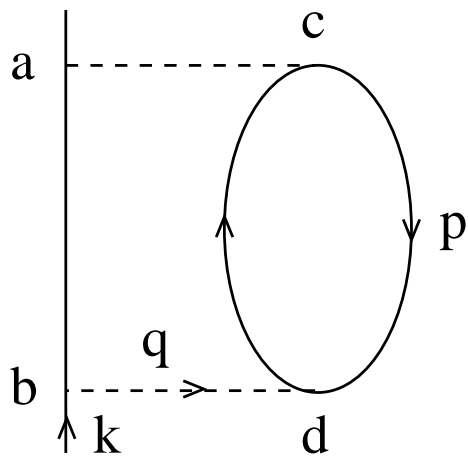
{\bf Fig.~1.} Generic Feynman diagram to evaluate the nucleon self-energy.
The indices $a,b,c,d$ are thermal indices. The nucleon propagator corresponding
to the baryonic lines is given in eqs. (5-9).
\medskip

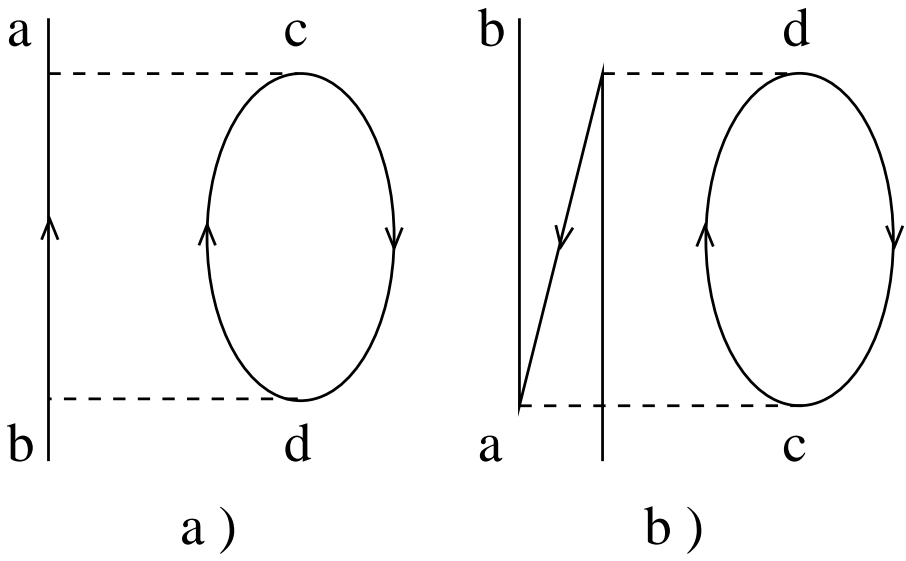
{\bf Fig.~2.} $a)$ polarization, $b)$ correlation graphs contributing to 
$Im\Sigma$ at $T = 0$ and contained in fig. 1. Here the direction of the arrows
stands for the conventional hole ( down ) and particle ( up ) propagators.
\medskip

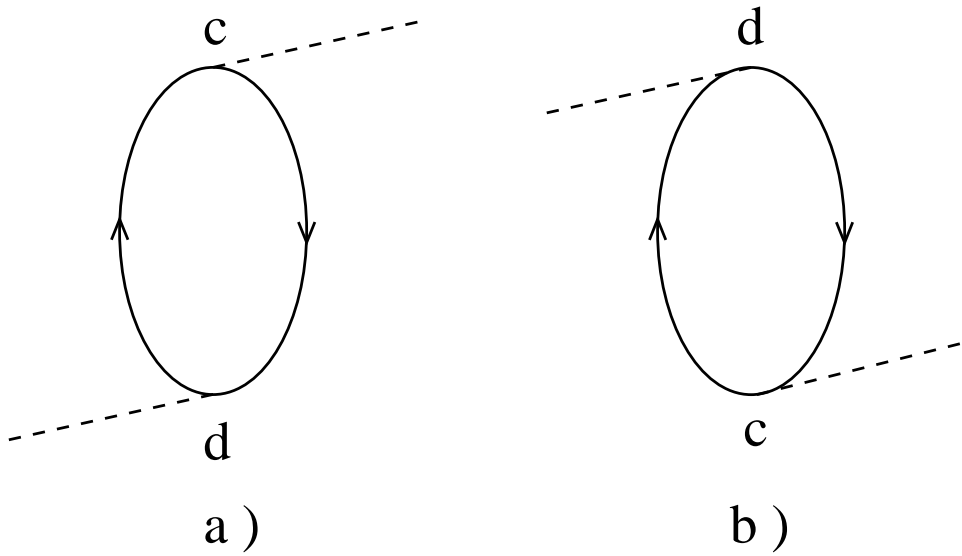
{\bf Fig.~3.} Polarization graphs appearing at $T = 0$ with the same notation
for the lines as in fig. 2.
\medskip

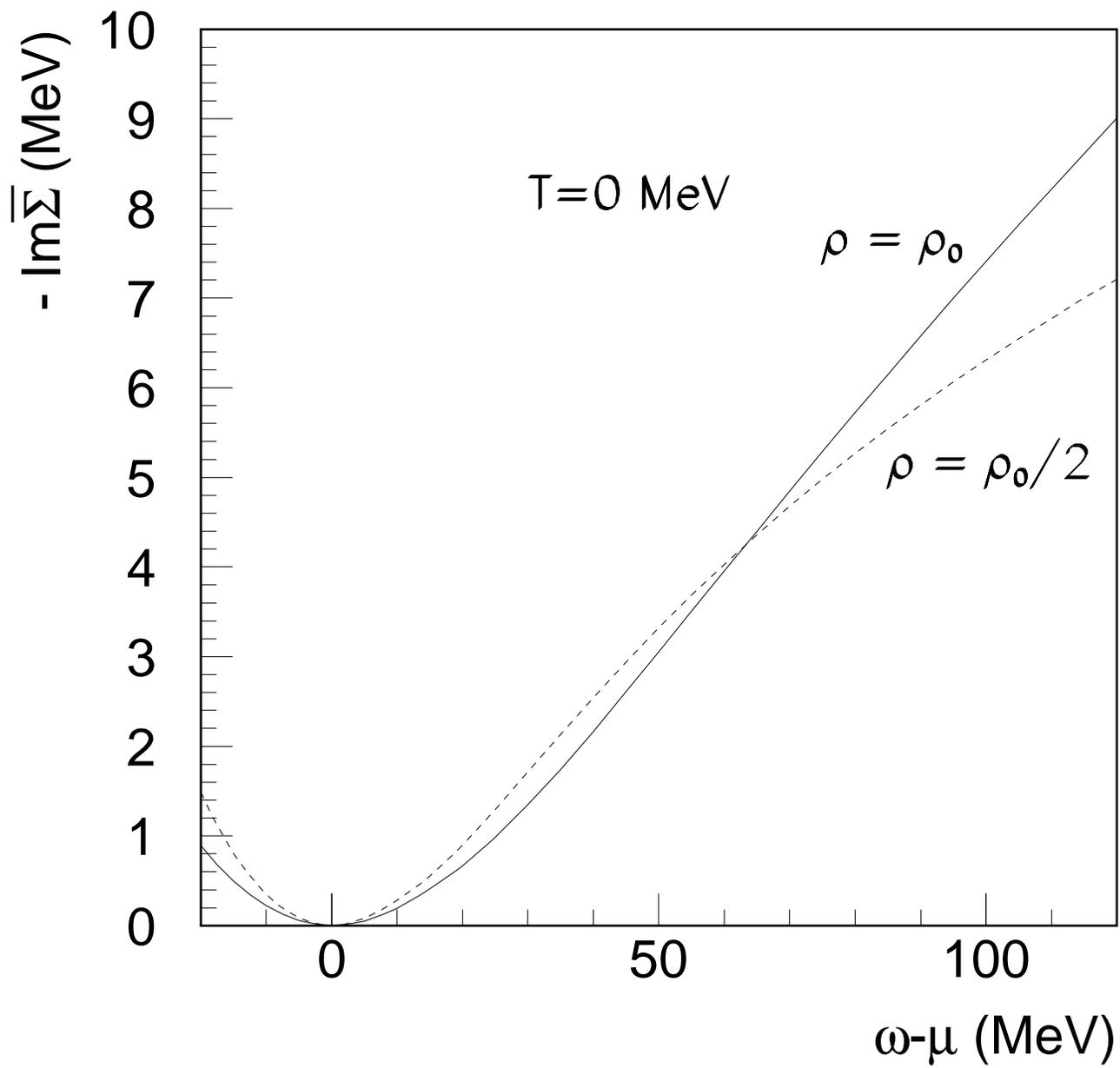
{\bf Fig.~4.} $- Im \bar{\Sigma} (\omega,k)$ at $T = 0$ as a function of 
$\omega-\mu$,
with $k = \sqrt{2 M \omega}$, evaluated for two densities.
\medskip

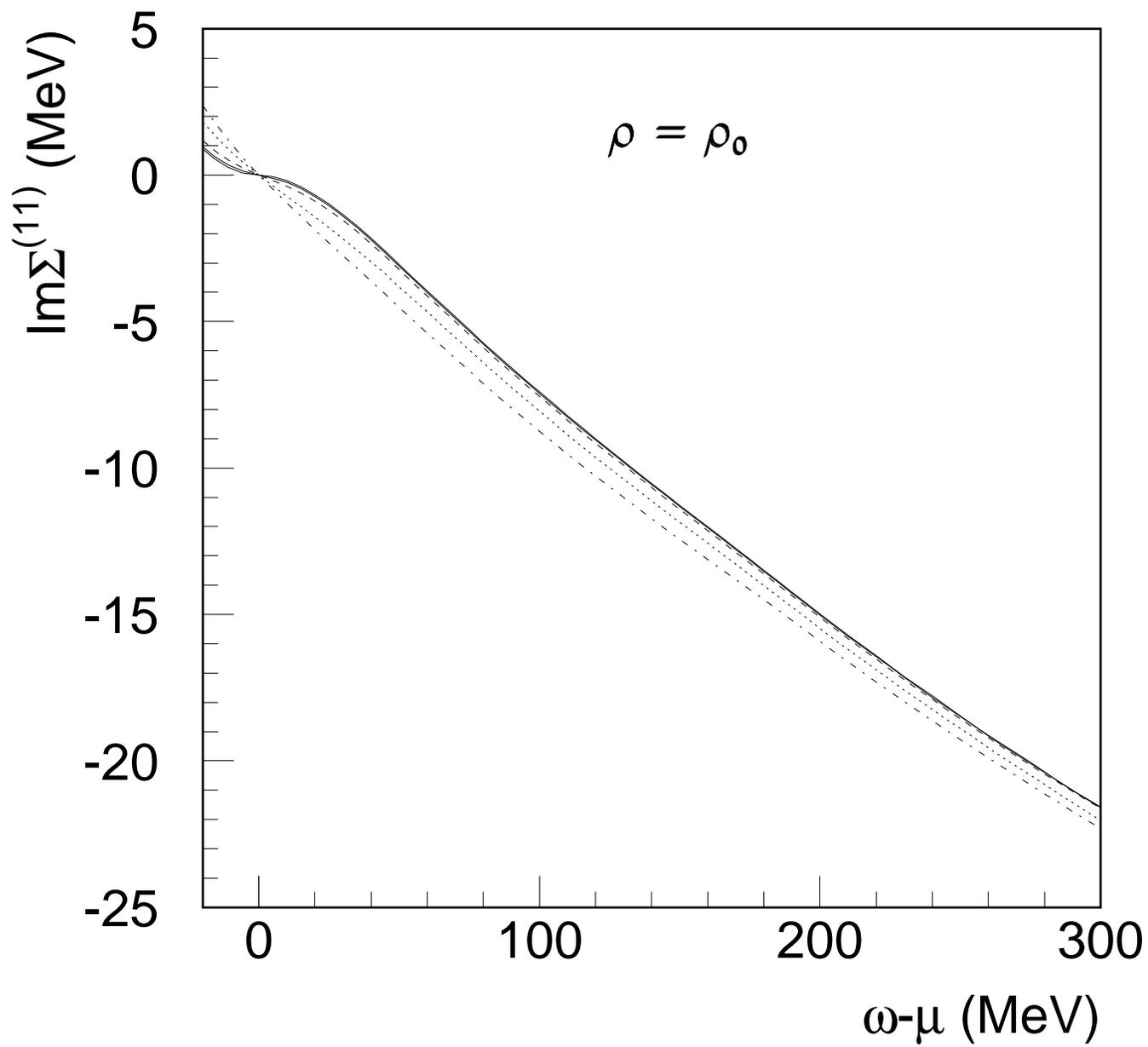
{\bf Fig.~5.} $Im \Sigma^{(11)} (\omega,k)$ at $\rho = \rho_{0}$ for several 
temperatures as a function of $\omega-\mu$ with $k = \sqrt{2 M \omega}$.
The solid line is for $T = 0$. The other curves correspond to $T = 2 \, MeV$ 
( long
dashed-dotted line ), $T = 4 \, MeV$ ( dashed line ), $T = 10 \, MeV$ 
( dotted 
line ) and $T = 20 \, MeV$  ( short dashed-dotted line ). At values 
$\omega-\mu <
0$, they appear correlatively with increasing values of $Im\Sigma^{(11)}$ as $T$
increases.
\medskip

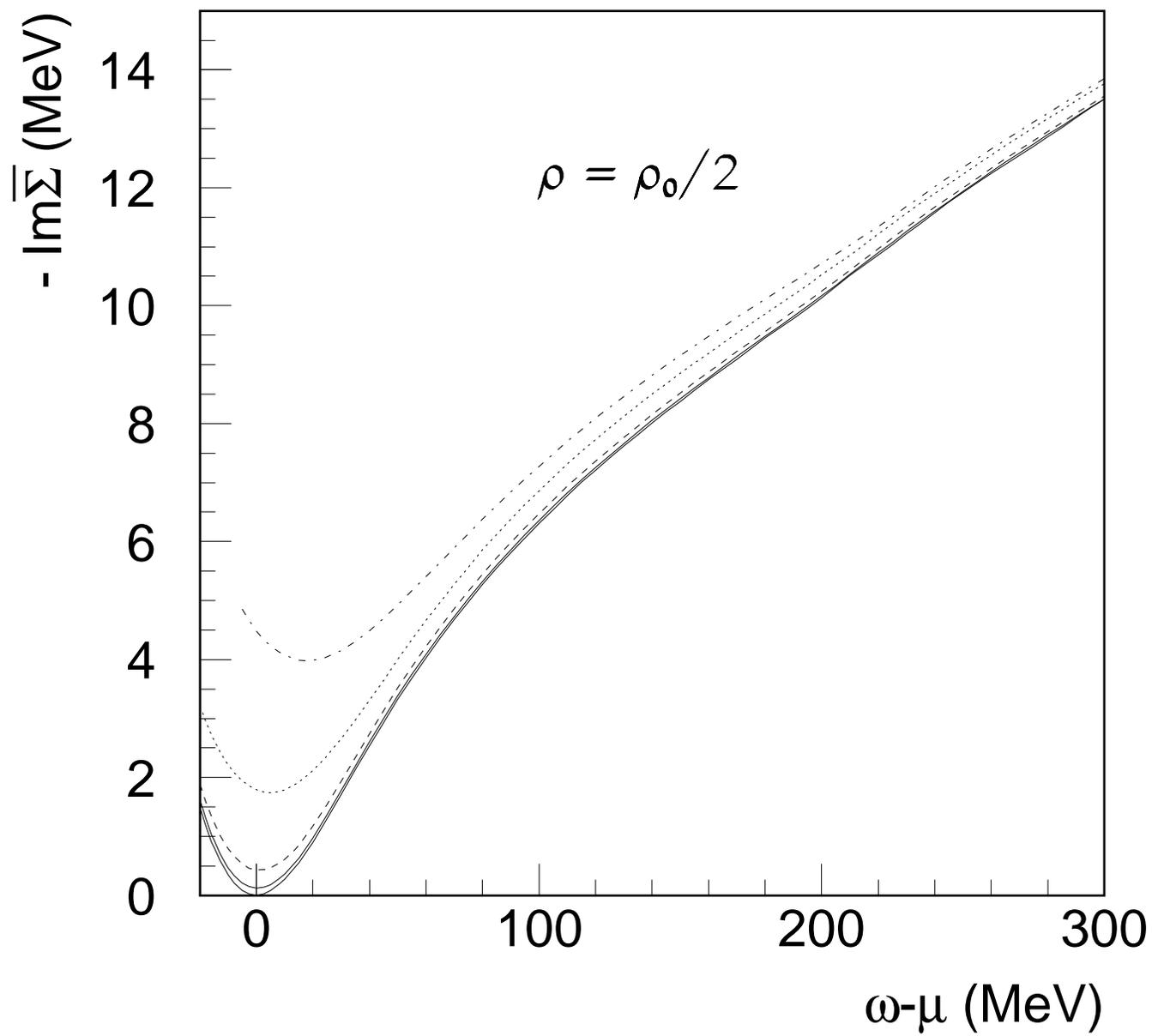
{\bf Fig.~6.} $- Im \bar{\Sigma} (\omega,k)$ at $\rho = \rho_{0}/2$ as a 
function of
$\omega - \mu$ for $k = \sqrt{2 M \omega}$ for several temperatures 
$T = 0, 2, 4,
 10, 20 \, MeV$ with the same notation as in fig. 5. $- Im \bar{\Sigma}$ 
increases with increasing $T$.
\medskip

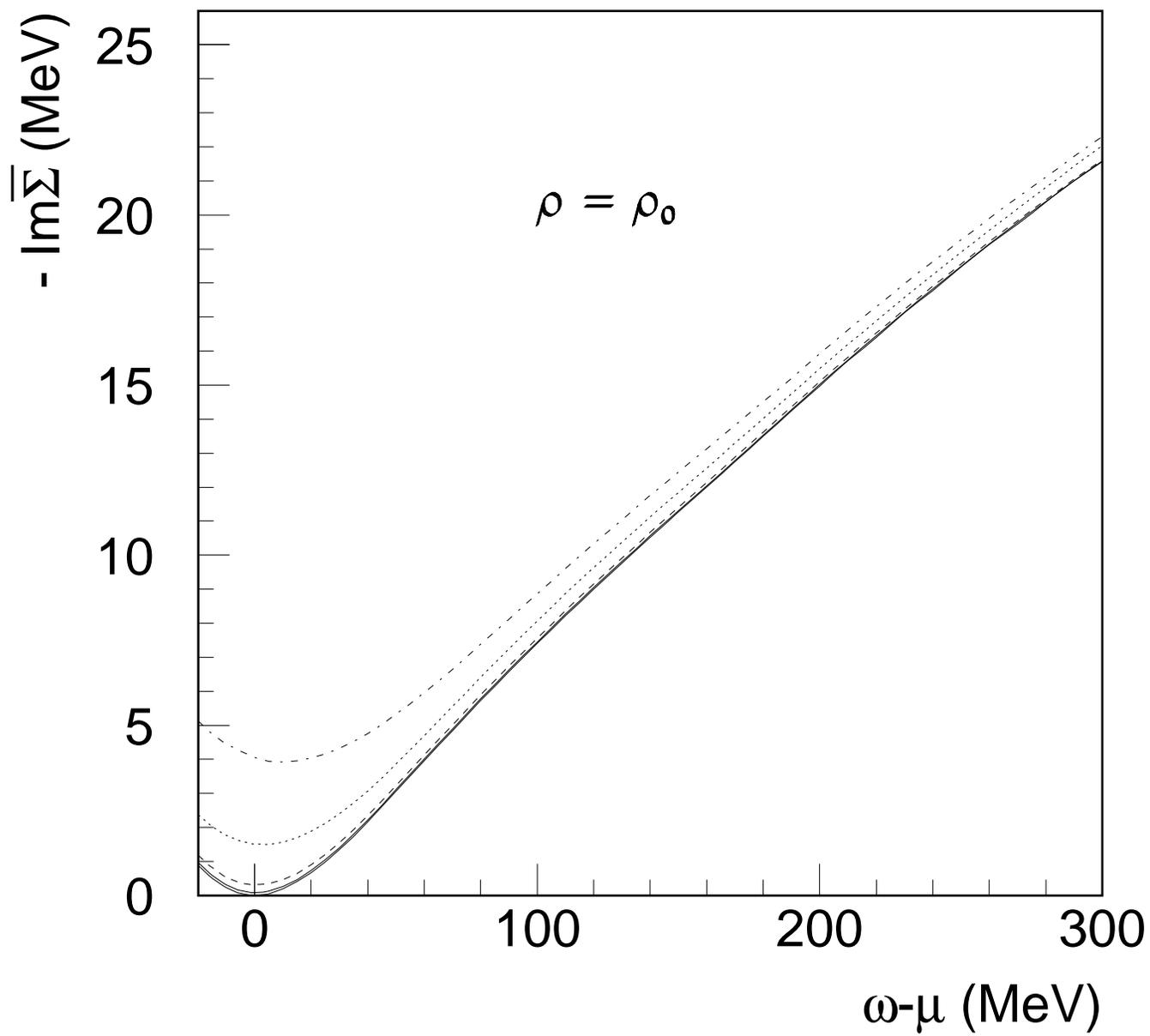
{\bf Fig.~7.}  Same as fig. 6 at $\rho = \rho_{0}$.
\medskip

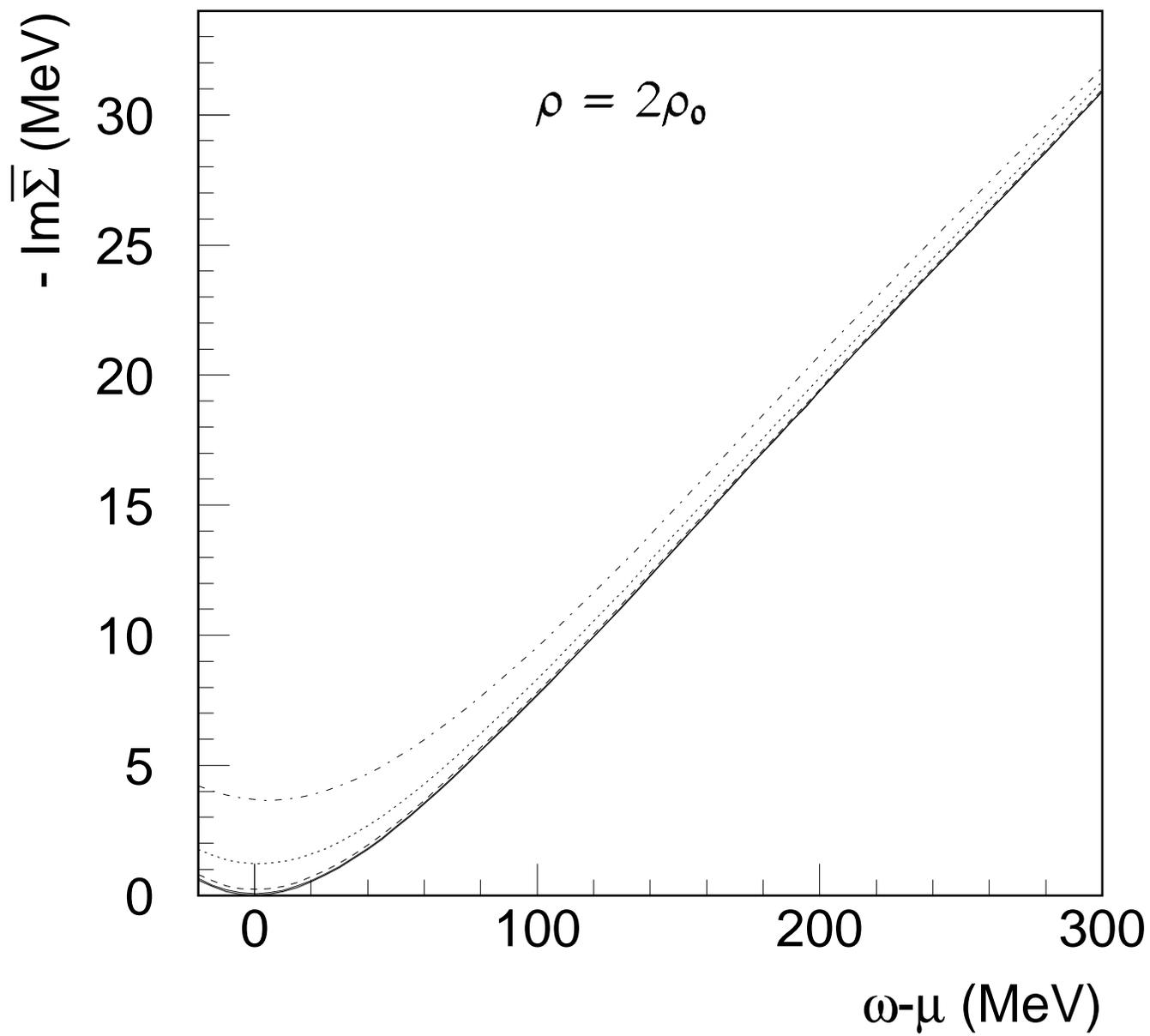
{\bf Fig.~8.}  Same as fig. 6 at $\rho = 2 \rho_{0}$.

\newpage

\begin{table}  
{\bf Table I.}
\vskip0.5cm

\begin{tabular}{|cc|c|c|c|c|c|}
\hline
 &T  & 0 MeV & 2 MeV & 4 MeV & 10 MeV & 20 MeV \\
$a_n$ & & & & & & \\
\hline
 $a_0$ & & -2.549  & -2.305 & -1.425 & 0.982 & 5.107  \\
 \hline
 $a_1$ & & 9.658 $10^{-2}$ & 9.098 $10^{-2}$ & 7.043 $10^{-2}$ & 
 3.404 $10^{-2}$ & -2.170 $10^{-2}$ \\
 \hline
 $a_2$ & & -9.706 $10^{-5}$ & -3.996 $10^{-5}$ & 1.569 $10^{-4}$ & 
 4.367 $10^{-4}$ & 8.413 $10^{-4}$ \\
 \hline
 $a_3$ & & -4.349 $10^{-7}$ & -6.674 $10^{-7}$ & -1.435 $10^{-6}$ &
 -2.408 $10^{-6}$ & -3.857 $10^{-6}$ \\
\hline
 $a_4$ & & 1.135 $10^{-9}$ & 1.456 $10^{-9}$ & 2.487 $10^{-9}$ &
 3.724 $10^{-9}$ & 5.709 $10^{-9}$ \\
\hline
\end{tabular}
\end{table}

\newpage

\begin{table}  
{\bf Table II.}
\vskip0.5cm

\begin{tabular}{|cc|c|c|c|c|c|}
\hline
 &T  & 0 MeV & 2 MeV & 4 MeV & 10 MeV & 20 MeV \\
$a_n$ & & & & & & \\
\hline
 $a_0$ & & 0.553 & 0.411 & 0.793 & 1.695 & 3.192 \\
 \hline
 $a_1$ & & -3.541 $10^{-2}$ & -2.722 $10^{-2}$ & -3.709 $10^{-2}$ & 
 -4.496 $10^{-2}$ & -4.430 $10^{-2}$ \\
 \hline
 $a_2$ & & 7.790 $10^{-4}$ & 6.834 $10^{-4}$ & 7.846 $10^{-4}$ &
 8.343 $10^{-4}$ & 7.542 $10^{-4}$ \\
 \hline
 $a_3$ & & -2.860 $10^{-6}$ & -2.458 $10^{-6}$ & -2.861 $10^{-6}$ &
 -3.027 $10^{-6}$ & -2.651 $10^{-6}$ \\
\hline
 $a_4$ & & 3.521 $10^{-9}$ & 2.963 $10^{-9}$ & 3.501 $10^{-9}$ &
 3.719 $10^{-9}$ & 3.215 $10^{-9}$ \\
\hline
\end{tabular}
\end{table}

\newpage

\begin{table}  
{\bf Table III.}
\vskip0.5cm

\begin{tabular}{|cc|c|c|c|c|c|}
\hline
 &T  & 0 MeV & 2 MeV & 4 MeV & 10 MeV & 20 MeV \\
$a_n$ & & & & & & \\
\hline
 $a_0$ & & 1.493 & 1.382 & 1.347 & 2.205 & 2.861 \\
 \hline
 $a_1$ & & -5.370 $10^{-2}$ & -5.017 $10^{-2}$ & -4.795 $10^{-2}$ & 
 -6.283 $10^{-2}$ & -6.065 $10^{-2}$  \\
 \hline
 $a_2$ & & 5.866 $10^{-4}$ & 5.555 $10^{-4}$ & 5.355 $10^{-4}$ &
 6.533 $10^{-4}$ & 6.214 $10^{-4}$ \\
 \hline
 $a_3$ & & -1.659 $10^{-6}$ & -1.553 $10^{-6}$ & -1.484 $10^{-6}$ &
 -1.874 $10^{-6}$ & -1.770 $10^{-6}$ \\
\hline
 $a_4$ & & 1.695 $10^{-9}$ & 1.570 $10^{-9}$ & 1.488 $10^{-9}$ &
 1.940 $10^{-9}$ & 1.834 $10^{-9}$ \\
\hline
\end{tabular}
\end{table}

\end{document}